# Highly degenerate photonic flat bands arising from complete graph configurations

Hanyu Wang[1†], Biao Yang[1,2†*], Wei Xu[1], Yuancheng Fan[2], Qinghua Guo[2], Zhihong Zhu[1*], C. T. Chan[2*]

1. College of Advanced Interdisciplinary Studies, National University of Defense Technology, Changsha 410073, China.
2. Department of Physics and Center for Metamaterials Research, The Hong Kong University of Science and Technology, Hong Kong, China.

*Correspondence to: yangbiaocam@nudt.edu.cn; zzhwcx@163.com; phchan@ust.hk

†These authors contributed equally to this work.

**Abstract**

**Inspired by complete graph theory, we demonstrate that a metallic claw "meta-atom" structure can carry a high number of nearly degenerate resonant modes. A photonic meta-crystal composing of a lattice of such meta-atoms exhibits a large number of flat bands that are squeezed into a narrow frequency window, and these flat bands can be designed to locate in a wide complete 3D bandgap. The degeneracy dimension ($N_f$) of the flat bands is determined by the number of branches ($N_b$) of the metallic claw with $N_f=N_b-3$, which is geometrically related to the complete graph theory. Different from those flat bands emerging from special lattice arrangements (e.g., Kagome lattice), the isolated flat bands here are**

**insensitive to lattice perturbations. The proposed mechanism offers a new platform for realizing various dispersion-less phenomena and a new paradigm to realize high density of states and spectra compressing.**

Introduction

Flat bands[1-5] refer to that spectral bands are dispersion-less or nearly so and their energy spectrum $E(\vec{k})$ are almost independent of momentum $\vec{k}$. In photonics, the realization of flat bands has been long pursued[6] for enhancing light-matter interaction with slow light[7-9] and wave localization[10,11], or it offers platforms for other applications such as distortion-free imaging and pulse buffering in nonlinear optics[12,13]. Typically, flat bands are found in the Dice[1], Lieb[3], Kagome[14,15] and other lattices[16,17] due to destructive interference, where fine-tuned nearest- and next-nearest neighboring hopping parameters are the key factors. Researchers have used waveguide arrays [10,18-23], dielectric/plasmonic resonators[24-32] and fine-tuned photonic crystals[33-35] to realize photonic flat bands, where high dielectric contrast or exact lattice symmetry are required. However, these mechanisms inspired by analogies with "frustrated" condensed matter systems[5,36,37] exhibit many limitations in the photonic regime where photonic bands usually arise from multiple coherent scattering rather than the hopping of local atomic orbits. Moreover, realization of full three dimension flat bands[38] is very challenging using such lattice arrangement.

In this letter, we demonstrate that three-dimensional (3D) flat bands can be realized using a periodic array of claw-like metallic structures. Isolated flat bands[39,40] existing in an absolute band gap emerge in such photonic crystals, and the flat-band modes are strongly localized near the meta-atoms. Different from the previous works, the flat bands are extremely stable when lattice constant changes. In addition, we find an interesting relation between the number of claw-branches and the number of flat bands, and the relation can be understood readily when we see that the claw-like structure is a manifestation of a "complete graph" from a geometrical point of view.

Recently, geometric aspects of physics have attracted a lot of attention. In photonics, various topological photonic semimetals and insulators[41] have been theoretically proposed and experimentally verified. Their properties are characterized by integers and are stable against local perturbations. For example, the number of edge states of photonic Haldane model[42,43] is directly related to its bulk integer Chern number defined in momentum (parameter) space[44]. Here, the real space geometry of the meta-atom generates interesting physics. We will see below that the electromagnetic response of the metallic claw structure can be described by a dynamic matrix that mimic the Laplacian matrix in complete graph theory[45], giving rise to highly degenerate localized modes which become the flat bands when the meta-atoms form a 3D crystal.

Figure 1a shows a simple complete graph $K_4$, where each pair of graph vertices (red dots) is connected by an edge (black line). There are $n$ vertices and $\frac{n(n-1)}{2}$ undirected edges in a complete graph $K_n$. Mathematically, each complete graph $K_n$ is precisely characterized by Laplacian matrix. For example,

$$L(K_4) = \begin{bmatrix} 3 & -1 & -1 & -1 \\ -1 & 3 & -1 & -1 \\ -1 & -1 & 3 & -1 \\ -1 & -1 & -1 & 3 \end{bmatrix} \qquad (1)$$

where each entry indicates the connection conditions, such as $L_{ii} = 3$ being the degree of vertex $i$ and $L_{ij} = -1$ when $i \neq j$. From a physics viewpoint, each entry of the matrix can also be read as the hopping term in a tight-binding model.

Generally, the spectra of Laplacian matrix are given by,

$$SpecL(K_n) = \left[0, \underbrace{n, \ldots, n}_{n-1}\right] \qquad (2)$$

which are highly degenerate except for the first 0-eigenvalue state with eigenvector $[1, \ldots, 1]_{1 \times n}$. The flat bands of our meta-crystal arise from the $n-1$ non-zero degenerate eigenvalues.

**Results: Metallic claw structure and photonic flat bands**

We start with a metallic claw structure with C$_4$ rotation symmetry. As shown in Fig. 1(b), each meta-atom consists of two perpendicularly placed split ring resonators (SRRs) which touch each other on the top, forming a four-branch claw. The whole meta-crystal is formed by arranging the claw structure in a primitive 3D tetragonal lattice. In the simulation, we assume the hosting material is air and regard metallic components as perfect electric conductors (PEC), which is a good approximation in microwave, teraherz and even far-infrared bands. Figure 1(c) schematically shows the effective electric circuit describing the electromagnetic response of the metallic claw structure, where each capacitor works as an edge in the complete graph and there are 6 possible edges. We have effectively introduced the capacitance $C'$ (between next-nearest branches) to distinguish with $C$ arising from those nearest neighbor branches.

The CST simulated photonic band structures and density of states (DOS)[46] are shown in Fig. 1(e, f) with the first Brillouin zone (FBZ) depicted in Fig. 1(d). The structural parameters are given in the figure caption. There is a 3D bandgap between the second and fourth bands, inside which lies a third band that is flat throughout the 3D FBZ and has no crossing with other bulk bands. We dubbed it an isolated flat band. The isolated flat band has almost vanishing dispersion among the full 3D FBZ, while the flat bands found in many previously works are typically dispersion-less in some particular planes or along some particular directions. As the flat band lines in a gap with the width of 1.73GHz at around 25 GHz, it cannot hybridize with the dispersive states and spans an orthogonal state-space by itself.

**Relationship with complete graph**

The flat bands originate from a set of highly localized degenerate modes of the meta-atoms. The underlying mechanism can be understood through local potential orbitals by analyzing the equivalent electric circuit consisting of capacitors and inductors (Fig. 1c). Following Chua's circuit notation[47,48], the Lagrangian of the circuit reads,

$$\mathcal{L} = \frac{C}{2}[(\dot{\varphi}_1 - \dot{\varphi}_2)^2 + (\dot{\varphi}_2 - \dot{\varphi}_3)^2 + (\dot{\varphi}_3 - \dot{\varphi}_4)^2 + (\dot{\varphi}_4 - \dot{\varphi}_1)^2]$$
$$+ \frac{C'}{2}[(\dot{\varphi}_1 - \dot{\varphi}_3)^2 + (\dot{\varphi}_2 - \dot{\varphi}_4)^2] - \frac{1}{2L}(\varphi_1^2 + \varphi_2^2 + \varphi_3^2 + \varphi_4^2)$$

(3)

where $\varphi_m$ indicates local potential on the branch $m$ as shown in Fig. 1(b-c). Without loss of generality, we have assumed $\varphi_0 = 0$. The Euler-Lagrange equation of motion is then,

$$I_1 = \frac{d}{dt}\frac{\partial \mathcal{L}}{\partial \dot{\varphi}_1} - \frac{\partial \mathcal{L}}{\partial \varphi_1} = C(2\ddot{\varphi}_1 - \ddot{\varphi}_2 - \ddot{\varphi}_4) + C'(\ddot{\varphi}_1 - \ddot{\varphi}_3) + \frac{1}{L}\varphi_1 \tag{4}$$

where $I_1$ indicating external current has been set to 0 as our system is source-free. In the same way, one can obtain $I_{2,3,4} = 0$ and expressing in the matrix form, we get,

$$(\omega^2 H - \frac{1}{L}I_{4\times 4})\Phi = 0 \tag{5}$$

where,

$$H = \begin{bmatrix} 2C+C' & -C & -C' & -C \\ -C & 2C+C' & -C & -C' \\ -C' & -C & 2C+C' & -C \\ -C & -C' & -C & 2C+C' \end{bmatrix} \tag{6}$$

and $\Phi^T = [\varphi_1, \varphi_2, \varphi_3, \varphi_4]$. The dynamic matrix $H$ shows strong resemblance to Laplacian matrix as mentioned above (Eq. 1). Different from the ideal complete graph $K_4$, there are two sets of weighted edges $C$ and $C'$ which are slightly different.

The condition of $\det(\omega^2 H - \frac{1}{L}I_{4\times 4}) = 0$ gives non-zero solution to Eq. 5 as,

$$\begin{cases} \omega_3 = \frac{1}{2\sqrt{CL}}, \Phi_3^T = \frac{1}{2}[-1,1,-1,1]e^{-i\omega t} \\ \omega_4 = \frac{1}{\sqrt{2(C+C')L}}, \Phi_4^T = \frac{1}{\sqrt{2}}[0,-1,0,1]e^{-i\omega t} \\ \omega_5 = \frac{1}{\sqrt{2(C+C')L}}, \Phi_5^T = \frac{1}{\sqrt{2}}[-1,0,1,0]e^{-i\omega t} \end{cases} \tag{7}$$

where $\Phi_4$ and $\Phi_5$ are degenerate states representing the two orthogonal dipole moments ($p_{x,y}$) of two independent SRRs. Here, the most interesting mode $\Phi_3$ shows the symmetry of an $d_{x^2-y^2}$ orbital and is orthogonal to the $p_{x,y}$ orbitals. The flat band of the photonic crystal is the Bloch state comprising this $\Phi_3$ mode. This mode arises only when the two SRRs touch each other. The CST simulated electric/magnetic field distributions of the flat band from several high-symmetry k-point as shown in Fig. 2 corroborates with our circuit prediction. At Γ, electromagnetic eigen-fields on the cutting plane $z = 0$ oscillate symmetrically. At $X$, $M$ and $A$, the electromagnetic eigenfields remain almost the same, which further indicates the flatness of the isolated band (derived from $\Phi_3$), and the mode profiles are very similar for different momentum $\vec{k}$.

Similar to the spectra of a Laplacian matrix (Eq. 2), there are 3 non-zero eigenvalues in claw structure shown in Fig. 1b, with two of them being the dipole modes ($p_{x,y}$), leaving one degree of freedom to contribute to the flat band. For claws with more branches, we can predict a general relation between the number of flat bands ($N_f$) and number of branches ($N_b$) as $N_f = N_b - 3$. The dimension of the flat-band sub-space gets bigger with an increasing number of branches, squeezing more and more flat bands into a narrow band of frequency. The zero-eigenvalue mode of a single meta-atom corresponds to $p_z$ orbital excitation with electric/magnetic field shown in Fig. S1[49].

Schematically, Figure 3 shows the local orbitals giving rise to the flat bands with number of branches increasing from $N_b = 3$ to 6. Without going through a tedious derivation of the Lagrangian (See for example, Eq. S1-S4)[49], all of them can be simply solved using the corresponding Laplacian matrix of a weighted complete graph as shown in the first column, where edges with different colours represent different weights. In realistic metallic claw structures, the different weights correspond to different capacitances between pairs of branches. The level of degeneracy of those flat bands is simply determined by the differences of those capacitances. Although it is hard to make the capacitances exactly the same in practice, a symmetrical design can make the differences smaller. In the second column, we explicitly provide the potential distributions of the dipole excitations where red and light-blue colours indicate respectively positive and negative potential distributions with the size representing the amplitude. The blue arrows show the dipole magnitude and directions defined as $\vec{p} = \sum_i r_i \varphi_i$. The third column shows higher-multipole excitations. Different from the dipole excitations, the higher-multipole orbitals consist of alternating positive and negative potentials, thus the sum of potential distributions vanishes as $\sum_i r_i \varphi_i = 0$. The dimension of higher-multipole orbitals determines the number of flat bands ($N_f$). For the $K_3$ case, there is no higher-multipole excitation, which agrees well with the geometric prediction of $N_f = N_b - 3$. From $K_4$ on, the number of higher-multipole modes increases linearly with the number of branches as indicated by the orange arrow in Fig. 3.

In order to verify those predictions, we show in Fig. 4 the evolution of the band structure as the number of branches changes. In the left inset, we show the metallic claw structures with the number of branches ranging from $N_b = 3$ to 6. These claws are arranged in a primitive tetragonal lattice to build the photonic meta-crystal. For simplicity, we also assume all claw-structures possesses $C_{N_b}$ rotation symmetry in each unit cell, although it is not compatible with the tetragonal Bravais lattice when $N_b \neq 4$. The orientation of the claw structure within the unit cell does not affect the existence of the flat bands. In the right column, the corresponding band structures and DOS[46] for different $N_b$ are shown. We have normalized the DOS[46] within the interval of 0 to 1 to reveal the contrast between the flat bands and dipolar pass bands. In order to clearly count the number of isolated flat bands and to check more details, different line styles are used and indicated in the figure. We find that the bands will become flatter when each unit cell gets bigger (keeping the size of claw-structure the same). Therefore, the small dispersions of the flat bands originate from the weak coupling of the localized modes between neighbouring meta-atoms. However, when the lattice constant gets too big, the gap will close as the Bragg scattering of the dipolar modes becomes too weak to sustain a complete gap. In that limit, the flat bands will intersect with the dipolar band manifold and no longer exist in a clean absolute gap.

As predicated above, we observe a $N_f = N_b - 3$ rule for the claw structures. When $N_b = 3$, there is no flat band even though the bandgap width is larger than 1.4GHz[50]. The DOS peak at the frequency corresponds to the lower boundary of the bandgap. Isolated flat bands emerge when $N_b > 3$, and all are confined inside a narrow frequency window (from 24.24 GHz to 24.53 GHz for the structural parameters specified in Fig. 1). The number of flat bands inside the band gap increases with the number of branches in the claw, and the corresponding DOS in the flat-band frequency window grows dramatically, which can potentially facilitate applications that requires a high photonic DOS. The $K_{20}$ case is shown in Fig. S2(a)-(b)[49] to further illustrate this phenomena. In particular, there are 17 flat bands in total as one can check in Fig. S2(c)[49], confirming again the $N_f = N_b - 3$ rule.

A unique feature of the metallic claw design, as the realization of a complete graph, is that the dimension of flat-band set depends on the number of branches and as such, we can arbitrarily enlarge the flat-band sub-space without shifting their operating frequency. As the spectral property of a complete graph is mainly determined by geometry and connectivity, the structural details of the meta-atoms are unimportant, and hence the flat bands and the related phenomena are robust even if the real samples deviate from the theoretical design (Fig. S3, [49]). On the other hand, their very high DOS and the high Q-factors of the high order orbitals render these complete-graph-inspired systems sensitive to environmental external fields, making them good platforms for information sensing.

**Conclusion**

In conclusion, we have designed claw-like metallic meta-atoms inspired by complete graph theory. We investigated photonic crystals composing of these meta-atoms and found isolated flat bands confined in a narrow frequency window. Different from previous works based on lattice geometry, the flat-band mode is insensitive to structural and lattice parameter perturbations. We found a simple relation governing the number of flat bands ($N_f$), which can easily be understood by mapping to the corresponding weighted complete graph. The highly degenerate flat bands and the associated high DOS persist even if the lattice (translation symmetry) is destroyed as the phenomena originate from internal degrees of freedom of the metallic claws.

**Acknowledgments**

This work is supported by National Natural Science Foundation of China (Grant No. 11674396). The work in Hong Kong is supported by Research Grants Council of Hong Kong (AoE/P-02/12).

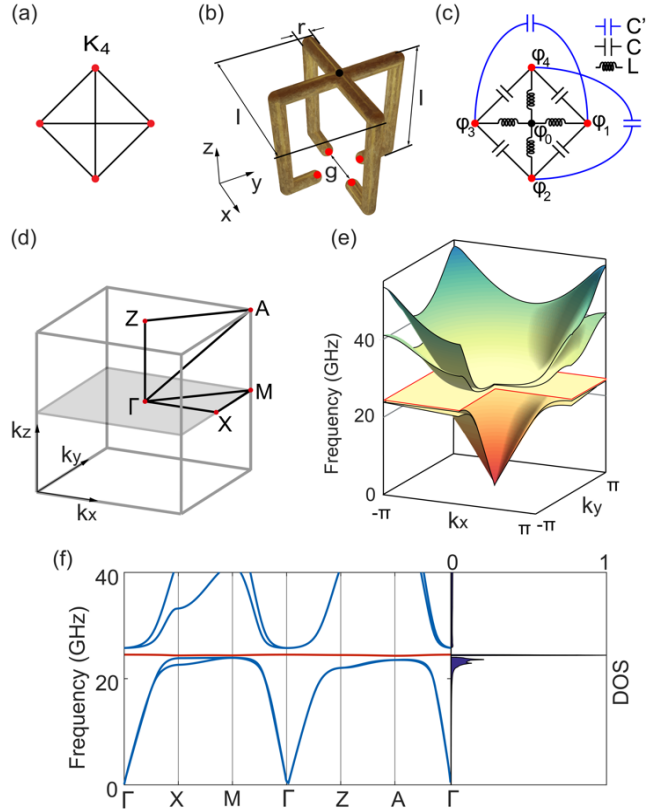

FIG. 1. Schematic view of metallic claw-like structures and flat bands. (a) Complete graph $K_4$ with four vertices and six edges. (b) Geometry of the claw-like metallic structure with $C_4$ rotation symmetry with length $l$=1.74mm, gap $g$=0.615mm, diameter $r$=0.21mm, as the building block of a primitive tetragonal photonic crystal with lattice constant $a$=3.5mm. (c) Effective electric circuit with inductors ($L$) and capacitors ($C$ and $C'$). (d) Three-dimensional first Brillioun zone (FBZ). Black solid lines show the paths followed by energy bands in (f). (e) Isolated flat band (surface with red edge) located in a complete band gap with $k_z = 0$. The fourth quadrant has been cut to show the sectional view for clarity. (f) Photonic band structure along a specific path as indicated in (d), with normalized density of states (DOS) shown in the right panel. The sharp peak in DOS stems from the flat band (red line, left panel) while zero DOS corresponds to the complete band gap.

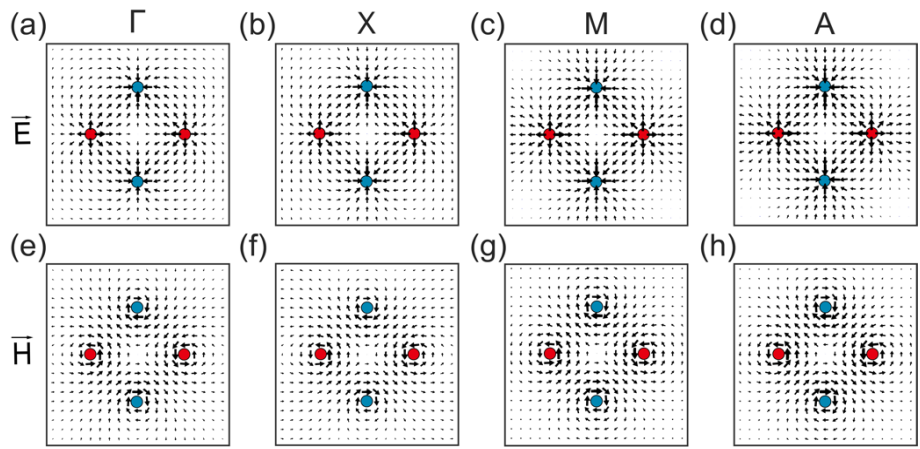

FIG. 2. Eigen-fields at high symmetry points of the isolated flat band for the $C_4$ metallic claw structure. The direction of arrow refers to the direction of electric (magnetic) field while its size representing the local intensity. The eigen-fields are in good agreement with our theoretical predication.

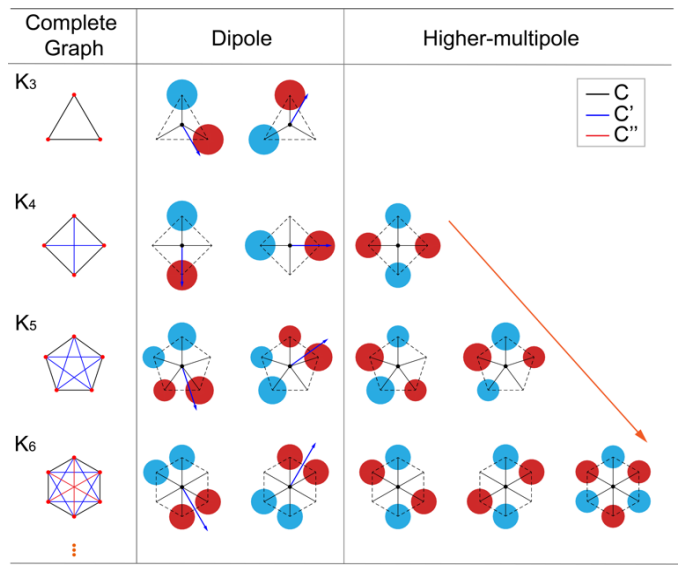

FIG. 3. Complete graph analogy of the metallic claw-like structures. The left column illustrates equivalent complete graph circuits of different $N_b$ (the number of branches for each "claw"), while the middle and the right columns show their corresponding eigen-potential distributions. Red (light-blue) disk represents positive (negative) electric potential, and its radius represents the absolute value of each potential. The number of eigenstates increases with $N_b$. For each single claw-like metallic structure, there are two non-collinear dipole excitations. Higher-multipole excitations appear when $N_b \geq 4$, concomitant with the emergence of flat bands. The number of flat bands ($N_f$) varies with $N_b$ as $N_f = N_b - 3$.

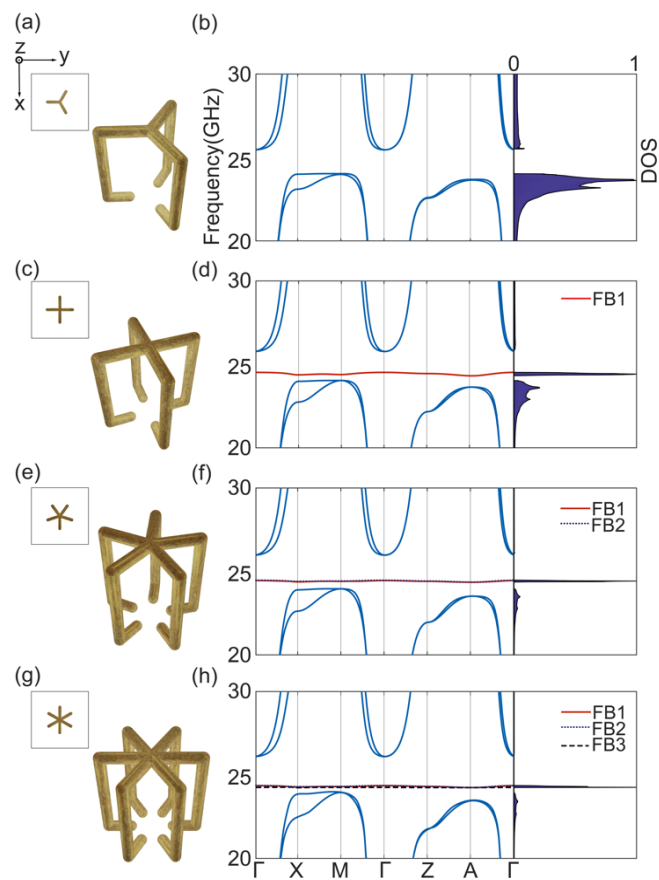

FIG. 4. The relation between number of branches $N_b$ and degeneracy of flat bands (FB) $N_f$. The left column (a, c, e and g) shows metallic claw structures with different $N_b$, with its top view presented in the top left inset. The right column (b, d, f and h) presents the corresponding band structure and DOS from 20GHz to 30 GHz. When $N_b$ increases, the shape of dispersive band remains almost unchanged. Furthermore, the flat bands remain at essentially the same frequency as $N_f$ increases, squeezing many bands into a narrow frequency window.

**Reference:**

[1] B. Sutherland, Physical Review B **34**, 5208 (1986).
[2] M. Arai, T. Tokihiro, T. Fujiwara, and M. Kohmoto, Physical Review B **38**, 1621 (1988).
[3] E. H. Lieb, Physical Review Letters **62**, 1201 (1989).
[4] A. Ramachandran, A. Andreanov, and S. Flach, Physical Review B **96**, 161104 (2017).
[5] D. Leykam, A. Andreanov, and S. Flach, Advances in Physics: X **3**, 1473052 (2018).
[6] D. Leykam and S. Flach, APL Photonics **3**, 070901 (2018).
[7] T. Baba, Nature Photonics **2**, 465 (2008).
[8] J. Li, T. P. White, L. O'Faolain, A. Gomez-Iglesias, and T. F. Krauss, Opt. Express **16**, 6227 (2008).
[9] S. A. Schulz, J. Upham, L. O'Faolain, and R. W. Boyd, Opt. Lett. **42**, 3243 (2017).
[10] S. Mukherjee, A. Spracklen, D. Choudhury, N. Goldman, P. Öhberg, E. Andersson, and R. R. Thomson, Physical Review Letters **114**, 245504 (2015).
[11] W. Maimaiti, S. Flach, and A. Andreanov, Physical Review B **99**, 125129 (2019).
[12] R. W. Boyd, J. Opt. Soc. Am. B **28**, A38 (2011).
[13] Z. Chen, M. Segev, and D. N. Christodoulides, Reports on Progress in Physics **75**, 086401 (2012).
[14] L. Santos, M. A. Baranov, J. I. Cirac, H. U. Everts, H. Fehrmann, and M. Lewenstein, Physical Review Letters **93**, 030601 (2004).
[15] A. Mielke, Journal of Physics A: Mathematical and General **25**, 4335 (1992).
[16] H. Tasaki, Phys Rev Lett **69**, 1608 (1992).
[17] H. Tasaki, The European Physical Journal B **64**, 365 (2008).
[18] A. Szameit and S. Nolte, Journal of Physics B: Atomic, Molecular and Optical Physics **43**, 163001 (2010).

[19] A. Crespi, G. Corrielli, G. D. Valle, R. Osellame, and S. Longhi, New Journal of Physics **15**, 013012 (2013).
[20] D. Guzmán-Silva, C. Mejía-Cortés, M. A. Bandres, M. C. Rechtsman, S. Weimann, S. Nolte, M. Segev, A. Szameit, and R. A. Vicencio, New Journal of Physics **16**, 063061 (2014).
[21] R. A. Vicencio, C. Cantillano, L. Morales-Inostroza, B. Real, C. Mejía-Cortés, S. Weimann, A. Szameit, and M. I. Molina, Physical Review Letters **114**, 245503 (2015).
[22] S. Xia, Y. Hu, D. Song, Y. Zong, L. Tang, and Z. Chen, Opt. Lett. **41**, 1435 (2016).
[23] S. Endo, T. Oka, and H. Aoki, Physical Review B **81**, 113104 (2010).
[24] F. Morichetti, C. Ferrari, A. Canciamilla, and A. Melloni, Laser & Photonics Reviews **6**, 74 (2012).
[25] Y. Nakata, T. Okada, T. Nakanishi, and M. Kitano, Physical Review B **85** (2012).
[26] M. Hafezi, S. Mittal, J. Fan, A. Migdall, and J. M. Taylor, Nature Photonics **7**, 1001 (2013).
[27] T. Jacqmin *et al.*, Physical Review Letters **112**, 116402 (2014).
[28] F. Baboux *et al.*, Physical Review Letters **116**, 066402 (2016).
[29] Y. Nakata, Y. Urade, T. Nakanishi, F. Miyamaru, M. W. Takeda, and M. Kitano, Physical Review A **93**, 043853 (2016).
[30] S. Kajiwara, Y. Urade, Y. Nakata, T. Nakanishi, and M. Kitano, Physical Review B **93**, 075126 (2016).
[31] S. Klembt, T. H. Harder, O. A. Egorov, K. Winkler, H. Suchomel, J. Beierlein, M. Emmerling, C. Schneider, and S. Höfling, Applied Physics Letters **111**, 231102 (2017).
[32] X.-Y. Zhu, S. K. Gupta, X.-C. Sun, C. He, G.-X. Li, J.-H. Jiang, X.-P. Liu, M.-H. Lu, and Y.-F. Chen, Opt. Express **26**, 24307 (2018).
[33] H. Takeda, T. Takashima, and K. Yoshino, Journal of Physics: Condensed Matter **16**, 6317 (2004).
[34] C. Xu, G. Wang, Z. H. Hang, J. Luo, C. T. Chan, and Y. Lai, Scientific Reports **5**, 18181 (2015).
[35] N. Myoung, H. C. Park, A. Ramachandran, E. Lidorikis, and J.-W. Ryu, Scientific Reports **9**, 2862 (2019).
[36] D. L. Bergman, C. Wu, and L. Balents, Physical Review B **78**, 125104 (2008).
[37] S. Flach, D. Leykam, J. D. Bodyfelt, P. Matthies, and A. S. Desyatnikov, EPL (Europhysics Letters) **105**, 30001 (2014).
[38] C. Weeks and M. Franz, Physical Review B **85**, 041104 (2012).
[39] D. Green, L. Santos, and C. Chamon, Physical Review B **82**, 075104 (2010).
[40] G. Tarnopolsky, A. J. Kruchkov, and A. Vishwanath, Physical Review Letters **122**, 106405 (2019).
[41] T. Ozawa *et al.*, Reviews of Modern Physics **91**, 015006 (2019).
[42] F. D. M. Haldane and S. Raghu, Physical Review Letters **100**, 013904 (2008).
[43] S. Raghu and F. D. M. Haldane, Physical Review A **78**, 033834 (2008).
[44] L. Lu, J. D. Joannopoulos, and M. Soljačić, Nature Photonics **8**, 821 (2014).
[45] J. A. B. a. U. S. R. Murty, *Graph theory with applications* (Macmillan, London, 1976), Accessed from https://nla.gov.au/nla.cat-vn2190177.
[46] B. Liu, S. G. Johnson, J. D. Joannopoulos, and L. Lu, Journal of Optics **20**, 044005 (2018).

[47] O. Feely, in *Chaos, CNN, Memristors and Beyond* (WORLD SCIENTIFIC, 2012), pp. 36.
[48] E. Zhao, Annals of Physics **399**, 289 (2018).
[49] See details in supplementary information.
[50] Y. Yang *et al.*, Nature **565**, 622 (2019).